# Accelerating Fast Fourier Transforms Using Hadoop® and CUDA®

Rostislav Tsiomenko
SAIC
Columbia, Md., USA
rostislav.tsiomenko@gmail.com

Bradley S. Rees[1]
SAIC
Columbia, Md., USA
bradley.s.rees@saic.com

*Abstract*— There has been considerable research into improving Fast Fourier Transform (FFT) performance through parallelization and optimization for specialized hardware. However, even with those advancements, processing of very large files, over 1TB in size, still remains prohibitively slow. Analysts performing signal processing are forced to wait hours or days for results, which results in a disruption of their workflow and a decrease in productivity. In this paper we present a unique approach that not only parallelizes the workload over multi-cores, but distributes the problem over a cluster of graphics processing unit (GPU)-equipped servers. By utilizing Hadoop® (The Apache Software Foundation) and CUDA® (NVIDIA Corporation), we can take advantage of inexpensive servers while still exceeding the processing power of a dedicated supercomputer, as demonstrated in our result using Amazon EC2® (Amazon Technologies, Inc.).

*Keywords*-Distributed computing; Parallel processing; Fast Fourier transforms

## I. INTRODUCTION

The Fast Fourier Transform (FFT) algorithm continues to play a critical role in many types of applications, from data compression, signal processing, and voice recognition, to image processing and simulation [5]. Our interest in the FFT algorithm relates to signal processing and its use in spectral analysis. Additionally, our interest lies in how to exploit existing FFT libraries to achieve high performance.

The Cooley–Tukey algorithm is the most commonly used FFT algorithm and has been refined and ported to a number of high performance platforms. Both the Math Kernel Library (MKL) from Intel Corporation [1] and the CUDA® FFT (CUFFT) library from NVIDIA Corporation [2] offer highly optimized variants of the Cooley-Tukey algorithm. NVIDIA claims that CUFFT offers up to a tenfold increase in performance over MKL when using the latest NVIDIA GPUs. However, even with the base $O(n \log n)$ time complexity of the Cooley-Tukey algorithm, and architecture-specific performance optimizations of MKL and CUDA, performing an FFT on large terabyte scale files remains computationally expensive.

In this paper, we present a novel approach for processing very large files that splits data processing across a cluster of servers, where each server could be equipped with GPUs for additional performance acceleration. This approach provides the ability to process very large files without the need of costly supercomputers. Runtime is estimated at $O(\frac{n \log n}{(0.8 * S) C})$ where C is the number of cores per server, S is the number of servers, and 0.8 is the efficiency factor (loss to overhead per server). We show that performance scales with the number of servers, with only a small loss of performance to Hadoop® overhead.

Our proof-of-concept software, discussed in Section III, uses Hadoop for handling file splitting and merging, and NVIDIA's CUDA for accelerating the computation of the FFT algorithm. We discuss results in section IV, where we baseline single server and single server with-GPU performance to illustrate the gains possible. We develop our conclusion in section V and present list of future research in section VI.

## II. RELATED WORK

The Cooley-Tukey algorithm is one of the most popular FFT algorithms, primarily due to its inherent support of parallelization. Chen and Gao [3] recently performed an analysis of the algorithm on multi-core servers to see if they could predict performance. Chen and Gao showed that performance scales with the number of cores in the

---

server and that they could predict the runtime with fairly good accuracy. That work could be extended to our case to predict the number of servers needed, assuming all servers have an equal number of cores.

In 1994 Agarwal, Gustavson, and Zubair [5] proposed a variant of the Argarwal-Cooley algorithm that is optimized for multi-core systems, specifically the IBM® SP (International Business Machines Corporation) platform. While their approach allowed the algorithm to scale to multiple cores, the algorithm requires a supercomputer class machine to achieve the throughput we sought, and is tied to the IBM platform. Additionally, the 1.25 GFLOP performance they reached in 1994, while impressive for the time, pales in comparison with what can be achieved currently with inexpensive NVIDIA GPUs that can reach ~375 GFLOPS. [2]

The performance benefit of using GPUs for FFTs has caught the attention of numerous researchers [4, 6, 7, 8, 9, 10]. However, the problem with GPUs is the limited amount of data that can fit within GPU memory. When the amount of data exceeds GPU memory, performance starts to degrade due to the latency of moving data between main memory and GPU memory. This problem was addressed by Gu, Siegel, and Li [6] who developed an efficient method for processing large data files between main memory and GPU memory.

Work closest to what we propose comes from Chen, Cui, and Mei [7] who developed an algorithm, along with supporting infrastructure, for performing FFTs using a GPU cluster. This work leverages a class of supercomputer that merges central processing units (CPUs) and GPUs for high-performance processing. GPU-based supercomputers ranked fifth, sixth, tenth, and fourteenth in the June 2012 release of the Top500 list of supercomputers [14].

Our goal is not to create a new variation of the FFT algorithm, or rely on expensive hardware platforms. Instead we would like to take advantage of the latest state-of-the-art implementation and be able to distribute the processing over multiple servers using Hadoop [12,13].

III. ALGORITHM DESCRIPTION

Our initial intuition was to split the data file into a number of Hadoop *Record*[2] objects encapsulating an *<offset, FFT segment>* pair, where *offset* indicates the byte offset from the beginning of the file, and *FFT segment* was an array of FFT-size samples (for an FFT size of 1024 and single-precision data, this would be 4096 bytes). However, this implementation was quickly deemed inefficient; given a 1TB input file, a 1024-point FFT, for example, would create 268,435,456 individual Records, with each Record holding only a 4096 byte value, thus requiring excessive overhead to track the completion progress of every Record inside the framework.

Besides creating unneeded complexity, this approach severely decreases the efficiency of CUDA; almost all GPU calculations occur faster than memory transfers from the host machine to the device (and vice versa) [15], as the on-GPU memory bus is much faster than the peripheral component interconnect express (PCI-e) bus, which has lower transfer speeds and higher latency. It is therefore beneficial to minimize memory transfers to and from the GPU.

In order to solve this problem, we implemented custom Hadoop input and output classes that treat a single Hadoop Distributed File System (HDFS) block as a *Split*[3] (the default Hadoop behavior), which is then treated as a single Record. The benefit of this approach is two-fold; using 512MB HDFS blocks, the amount of map tasks launched for a 1TB file decreases to 2,048, with each map task having 512MB of data to work with. This data can be transferred to the GPU in a single pair of allocate and memory copy operations, and the partitioning of FFT segments can be done inside memory using CUFFT's batched FFT plan.

The CPU implementation uses the same classes for file input and output, using a for-loop to go through the 512MB of data in-place. The block size was set at 512MB, as that was the maximum limit for pointers to in-place FFT data on the CUDA cards we used for development; however, this number can be easily adjusted by changing the Hadoop

---
[2] Record refers to a Hadoop class

[3] A Split is part of Hadoop's process of break input data into a number of smaller chunks.

*dfs.block.size* variable when copying the input file into HDFS, making it easy to change CUDA buffer sizes for clusters with different GPU configurations without rewriting the code. The default HDFS block size is 64MB; it is not clear if there are any Hadoop performance penalties for increasing this to 512MB or beyond, as the use of Hadoop block sizes larger than 128MB is poorly documented.

The traditional Reduce portion is problematic, however, as MapReduce will output as many files as there are reducers. Since we desire a single output file, one reducer has to be used; this results in all the data from our potentially terabyte-sized file being sent over the network to a single node, where it can be written back to HDFS. Thus, using this method would impose a significant penalty on the processing time, as one machine would have to receive all the map outputs from the network, and then write them back to the cluster. A work-around was devised where the number of reducers is set to 0, and each map task writes its output directly to HDFS. This eliminates the reduce phase, and the user must instead use a post operation –*getmerge* call to the HDFS, which merges files in the output directory (named by their position in the original file, and thus correctly sorted) and copies them to the local file system, cutting down on the time needed to obtain a local copy of the final output. This approach is bottlenecked by the speed of the system writing the final merged file to local disk. Figure 1 represents the workflow of our final Hadoop implementation.

IV. RESULTS AND ANALYSIS

In order to model the potential performance improvement from parallelization we utilize Amdahl's Argument, which states the maximum speedup is determined by

$$S(N) = \frac{1}{(1-P) + \frac{P}{N}},$$

where P is the proportion of calculations that can be parallelized, (1-P) is the proportion that cannot be parallelized, and N is the number of concurrent threads. Simply put, Amdahl's Argument states that parallelization can only speed up an algorithm to a certain point; in our case, the non-parallel portion of the program, in a single machine set-up, is the reading and writing of data to disk, thus making it the primary constraining factor for performance. The portion of the algorithm that is parallelized through the use of a GPU is the calculation of the FFT blocks; by moving the entire process to Hadoop, it is in theory possible to parallelize the reading and writing process as well.

In order to estimate the P and (1-P) proportions of the algorithm, we first ran single CPU and single GPU tests. File size was kept at 16GB to make multiple runs feasible. All single-machine tests were performed on an Intel® E6400 (two cores with hyper-threading enabled), NVIDIA® GT620 (96 CUDA cores) machine, utilizing JCUFFT (a Java® (Oracle America, Inc.) wrapper around CUFFT) and JTransforms (a native Java FFT library comparable

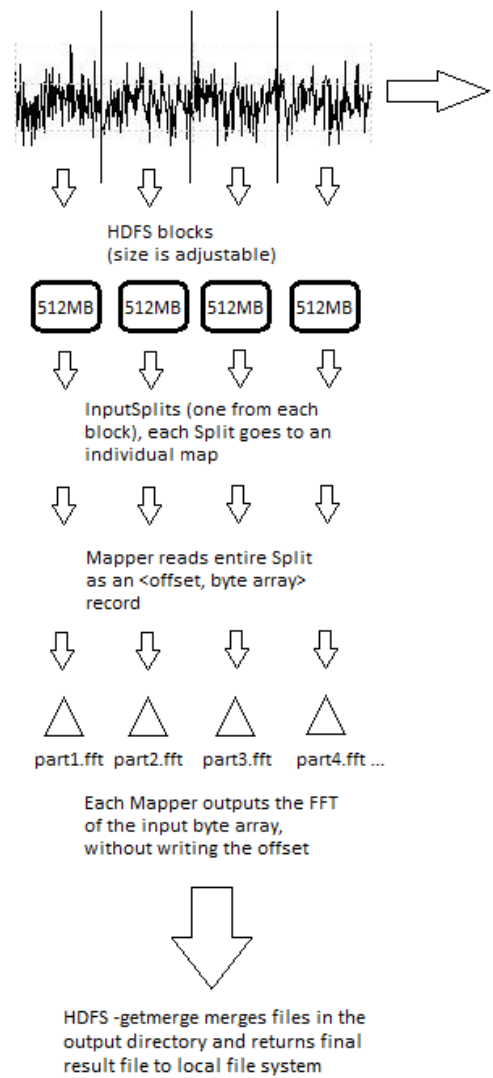

Figure 1: Hadoop Implementation

in performance to Intel's MKL):

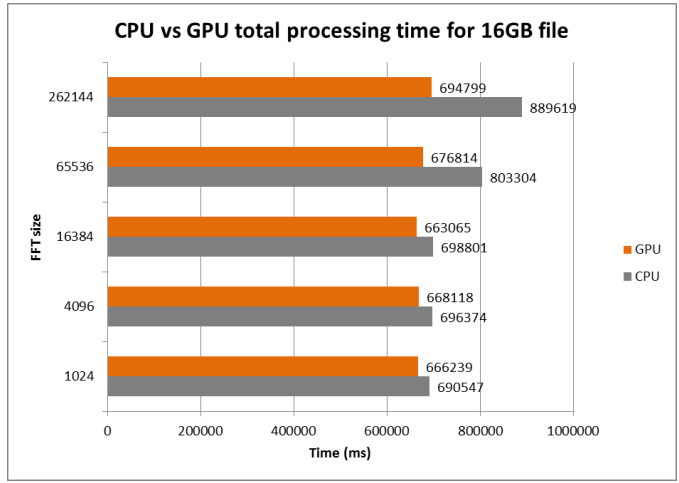

Figure 2: Total processing time for a 16GB file

Figure 2 shows that total processing time for the GPU implementation was lower only by 10% -15% on average. Figure 3 eliminates reading/writing from the benchmark times and only compares FFT calculation time:

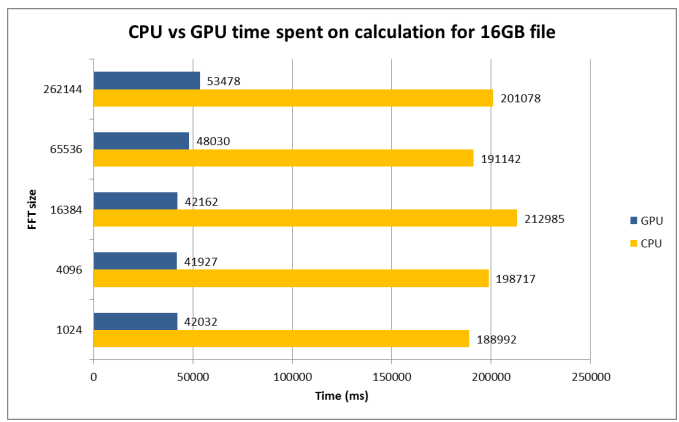

Figure 3: Total FFT calculation time

Figure 3 shows that NVIDIA's claim of a tenfold speedup is realizable, as we were able to decrease the total FFT calculation time by a factor of 5 on average using a 96-core GPU. The current NVIDIA flagship GPUs have more than 1500 cores, to put things into perspective. It follows logically that a major portion of the total calculation time must be taken up by input/output (I/O); which we illustrate in Figures 4 and 5:

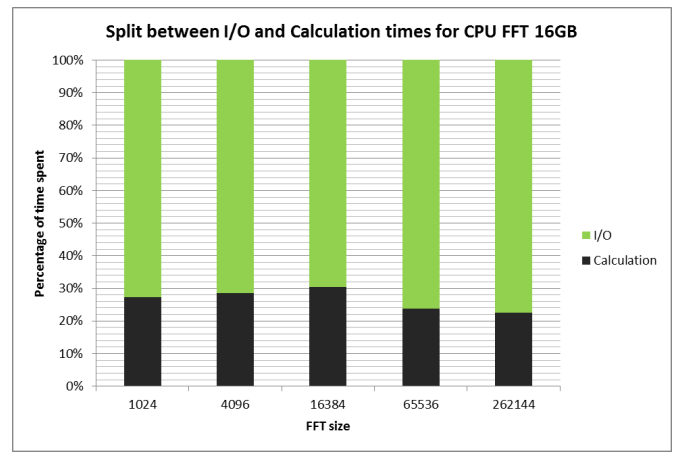

Figure 4: Percent of time spent in I/O and FFT calculation for CPU test

Benchmarking I/O and FFT calculation times separately showed that the CPU implementation spends an average of 70%-75% of total calculation time reading and writing from and to the disk; the remaining 20%-25% time calculating FFTs, even if reduced by a factor of 10, will not significantly decrease total calculation time (as seen in Figure 2). This is proven further in Figure 5, where the FFT calculation time for the GPU makes up only 5%-8% of total time spent processing the file, and I/O dominates the benchmark.

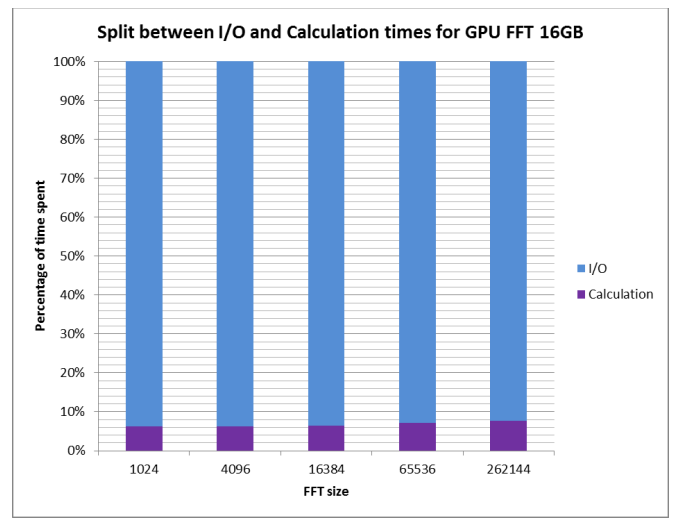

Figure 5: Percent of time spent in I/O and FFT calculation for GPU test

From these tests, we can deduce that reading and writing constitutes a large percentage of computation time for FFTs on large files, and though using a GPU is helpful, it can only take us so far. Although it is tempting to set P at 0.75, these results will vary quite a bit across different hardware platforms and FFT libraries; everything from number of CPU/GPU cores to hard disk read/write speed can change the variables in Amdahl's equation.

Measuring these differences becomes even harder when utilizing Hadoop; Even assuming the Hadoop cluster hardware is heterogeneous, other factors such as current network load, link speed, and varying background processes running on computation nodes may affect results.

In addition to the previously mentioned varying environment variables, obtaining a GPU cluster of an applicable size (at least 8 to 16 nodes) is expensive. We therefore decided to utilize Amazon's EC2® (Amazon Technologies, Inc.) cloud computing service, which offers GPU clusters for high performance computing (HPC) applications for hourly rental. Unfortunately we were only able to run on an eight-node cluster with limited benchmarking, the result of which can be seen in Figure 6.

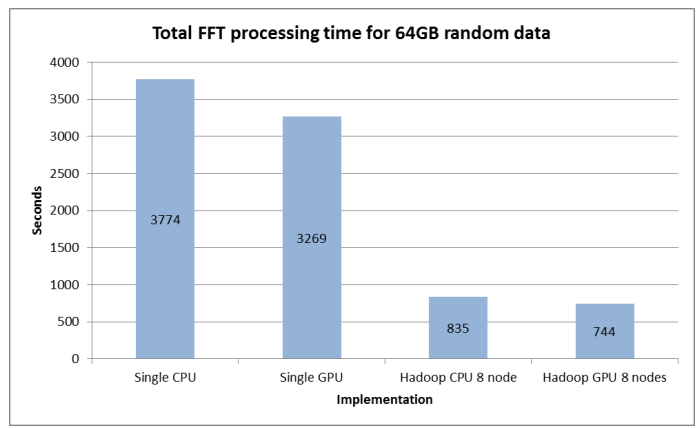

Figure 6: Single machine vs Hadoop EC2 computation times

## V. CONCLUSION

Hadoop is traditionally used for searching very large data sets, but that technology can be adapted to a wide variety of problems. In this paper, we have shown how Hadoop and GPUs could be combined to perform accelerated and distributed FFTs. Additionally, we have evidence to conclude that the combination can make FFT operation considerably faster than a single machine when processing terabyte size data files. This performance gain will reduce the wait time experienced by the signal analyst. Additionally, we have shown that the accelerated FFT performance can be obtained using commodity servers with inexpensive GPUs, including the use of Amazon's EC2 environment.

## VI. FUTURE WORK

Although the Hadoop implementation has tangible benefits, there is the possibility that performance could actually be better than what was observed. The Amazon's EC2 virtualized environment is not the optimal test bed for benchmarking, as the virtual instances launched by the user reside on non-dedicated hardware that is always in use by other virtual instances. Amazon does not provide a way to track CPU/GPU utilization for hardware, therefore the hosts (CPU/GPU) are always under unpredictable load. Furthermore, storage is entirely virtualized as well, and usually resides in a separate location from the virtual instance hosts, making it prone to fluctuations in network traffic and usage as well. In separate tests for I/O performance, we determined read/write speeds could vary by as much as 200%; using a virtualized operating system (OS) for Java benchmarking (by way of System.nanoTime() call) also leads to granularity issues, and it is not clear if a specialized benchmarking package would be able to fix this [16]. Additionally, Apache recommends against using Hadoop in a virtualized environment; the framework was designed for a cluster of dedicated commodity machines, and virtualization undermines some design concepts central to Hadoop [17].

We plan on performing additional tests on a dedicated cluster to determine optimal performance gains. Additionally, we plan on making a slight modification to the code that will allow for hybrid CPU/GPU clusters, which could compete with pure GPU clusters in terms of cost. CUDA 5 will contain support for RDMA (Remote Direct Memory Access) when released, meaning the concept of paralleling an FFT across a server cluster can be implemented much faster using C for both computation and network distribution, entirely eliminating Java and Hadoop from the process. Such an implementation, however, would be both expensive and time consuming.

Lastly, our first implementation does not support all types of FFT operations. We plan on expanding our work to allow overlapping FFTs operation to be performed in our distributed environment.

## VII. ACKNOLEDGEMENT

The authors would like to thank Benjamin Greenberg for his help in running the benchmarks, and Graham Cruickshank for his proof reading skill and support.